\documentstyle[aps]{revtex}

\newcommand{\pcite}[1]{{\cite{#1}}}
\newcommand{\scite}[1]{{\cite{#1}}}
\def\gs{\mathrel{\raise1.16pt\hbox{$>$}\kern-7.0pt 
\lower3.06pt\hbox{{$\scriptstyle \sim$}}}}         
\def\ls{\mathrel{\raise1.16pt\hbox{$<$}\kern-7.0pt 
\lower3.06pt\hbox{{$\scriptstyle \sim$}}}}         

\begin{document}
\bibliographystyle{apsrev}

\title{Detection limits for super-Hubble suppression of causal fluctuations}

\author{Arjun Berera\thanks{email:ab@ph.ed.ac.uk, PPARC Advanced Fellow}
and 
Alan F. Heavens\thanks{email: afh@roe.ac.uk}}
\address{$^{*}$ Department of Physics and Astronomy, University of
Edinburgh, King's Buildings, Mayfield Rd, Edinburgh EH9 3JZ,
U.K.\\ $^\dag$ Institute for Astronomy, University of Edinburgh, Royal
Observatory, Blackford Hill, Edinburgh EH9 3HJ, U.K.}
\maketitle


\begin{abstract}

We investigate to what extent future microwave background experiments
might be able to detect a suppression of fluctuation power on large
scales in flat and open universe models.  Such suppression would arise if
fluctuations are generated by causal processes, and a measurement of a
small suppression scale would be problematic for inflation models, but
consistent with many defect models.  More speculatively, a measurement
of a suppression scale of the order of the present Hubble radius could
provide independent evidence for a fine-tuned inflation model leading
to a low-density universe.  We find that, depending on the primordial
power spectrum, a suppression scale modestly larger than the visible
Horizon can be detected, but that the detectability drops very rapidly
with increasing scale. { }For models with two periods of inflation,
there is essentially no possibility of detecting a causal suppression
scale.

PACS No.: 98.80.Es, 98.70.Vc, 98.80.Cq

\end{abstract}

\medskip

In press Physical Review D 2000
 
\medskip

\maketitle

\section{Introduction}
\label{intro}

If we assume that fluctuations in the Universe were created by causal
processes, then an unambiguous feature of the power spectrum is the
maximum scale over which such fluctuations are correlated.  In
principle this scale is a characteristic of all density fields in the
Universe, matter, cosmic microwave background radiation (CMB) scalar
and tensor fluctuations, primordial magnetic fields etc.  However, the
effects of this scale are generally difficult to detect, as they
inevitably occur on a very large scale. Indeed, in inflationary
models, the scale may exceed the Hubble length $1/H$, and we will
refer to this scale as the super-Hubble suppression scale.

Efforts to understand the super-Hubble scale primarily have focused on
the CMB scalar perturbation power spectrum.  {}For scalar perturbations,
as well as any matter perturbations, if they are created by local
causal processes, it is well known that causality requires the power
spectrum to diminish at least as fast as $k^4$ at small $k$.  The
argument for this was first put forth by \scite{zel65}
and over the years it has been refined \cite{Pee74,CarrSilk83,at86,rw96}.

The advent of inflationary cosmology has placed specific focus on the
size of the super-Hubble scale, in particular relative to the Hubble
distance $1/H$ (we set $c=1$ throughout).  Whereas the Hubble distance
characterizes the maximum distance over which causal interactions can
occur during one expansion time, in distinction, the super-Hubble
length scale $\lambda_H$ characterizes the absolute largest scale over
the lifetime of the Universe, on which causal interactions can seed
density perturbations.  In particular the super-Hubble scale accounts
for the growth of all local perturbations since their conception up to
the present due to the expansion of the universe.  {}For example, in a
flat universe, a perturbation of the maximum causal physical length
created at time $t_i$ expands to a physical scale at the present time
$t_0$
\begin{equation}
\lambda(t_0; t_i) = \frac{R(t_0)}{R(t_i)} \int_{0}^{t_i}
\frac{dt'}{R(t')},
\end{equation}
where $R(t)$ is the cosmic scale factor.  Depending on the behaviour
of the cosmological scale factor $R(t)$, this can exceed the present
Hubble radius $1/H_0$.  In the regime of Standard Hot Big Bang
cosmology $R(t) \sim t^{1/2}$ and $t^{2/3}$ in the radiation- and
matter-dominated regimes respectively.  On the other hand, inflation
is a regime in which the Hubble parameter $H \equiv {\dot
R(t)}/{R(t)}$ is approximately constant and the scale factor is
accelerating, ${\ddot R(t)} > 0$, with generically a quasi-exponential
growth.  {}For example, if at $t_i$ during inflation a local
perturbation is created at scale $1/H$, by the end of inflation at
$t_f$ it will imply a super-Hubble scale of $\lambda(t_f) \approx
(1/H) \exp[H(t_f-t_i)]$. Thus immediately after even a few
$e-$foldings of inflation, $\lambda \gg 1/H$.

Outside the inflation regime, the causal horizon in the standard
Big-Bang regime grows in physical units as $\lambda \sim t$.  In
principle this horizon eventually will grow bigger than the one
set by inflation, which in physical units grows as $\lambda R(t)/R(t_f)
\sim t^a \lambda$ where $a\sim 1/2$ during
radiation-domination and $a\sim 2/3$ during matter-domination.
However the generic prediction of most inflation models is that
$\lambda(t_0) \gg \eta_0 \sim 2/H_0$, where $\eta_0$ is the physical
distance from last scattering or equivalently the causal horizon
for events generated since last scattering.

Thus, the generic expectation
for density perturbations produced solely
within non-inflationary regimes is $\lambda \sim 1/H_0$,
whereas isentropic inflation models \cite{Guth81,AS82,Linde82,Linde83,Olive90}
generally predict $\lambda \gg 1/H_0$ and non-isentropic
inflation models \cite{Berera95,Berera97} are fairly impartial
to any particular scale between these two limits.
As such, the super-Hubble scale, in principle, is an ideal
parameter for discriminating between inflationary and non-inflationary
models of density perturbations.  However the practical limitation is
that the super-Hubble scale can be measured accurately only if it no
greater than about $1/H_0$.  Thus at best one hopes to distinguish
from the data whether 
$\lambda \stackrel{<}{\sim} 1/H_0$ or 
$\lambda \stackrel{>}{\sim} 1/H_0$.  If
such a measurement could be reliably made, it would be far-reaching:
evidence for a maximum causal scale which was smaller
or of order the Hubble radius
would not be expected in inflation, but could be produced by, say,
topological defect models generating fluctuations causally after the
Big Bang.  A more speculative possibility is that inflation produced
only just enough e-foldings to solve the horizon problem, but not
enough to drive the present Universe to flatness.  The fine-tuning
required for this is problematical, but measurement of a super-Hubble
scale modestly larger than $1/H_0$ could provide evidence, independent 
of a measurement of $\Omega_0$, for such non-flat inflation.

{}For the Cosmic Background Explorer experiment (COBE), a
likelihood fit to the four-year data with respect to the index,
amplitude and super-Hubble suppression scale was done in
\scite{bfh97}.  Their results interestingly preferred a super-Hubble
suppression scale $\lambda \sim 4/H_0$ (or $k_{min}\eta_0 \sim 3.5$).
Since the preference was within one-sigma of an infinite suppression
scale, the conclusion of their analysis was that no super-Hubble
suppression scale $\lambda > 1/H_0$ was excluded.  However, the fact
that this analysis did not give a clear-cut measurement of an infinite
suppression scale leaves some margin of doubt.  Due to the importance
that detection of a small suppression scale would have in
discriminating cosmological models, it is advantageous at this stage
to explore the possibilities for measuring this scale. 

What are the
possibilities of improving on this measurement with new experiments?
The first point is that since the signal for super-Hubble suppression
is confined to low-order multipoles in the CMB signal, instrumental
noise is not the dominant error, and the higher signal-to-noise of MAP
and Planck will give little advantage over COBE.  However, there are
still two advantages. One is that the CMB power spectrum at low
multipoles is determined not just by the suppression scale, but other
parameters as well: the newer experiments will determine the other
parameters rather precisely, thus reducing the marginal error on the
suppression scale. Second, the better frequency coverage of the
later experiments (especially Planck) should allow better control of
systematic errors, especially through effective removal of foreground
sources.  For the quadrupole, the successful removal of as much
foreground emission from the Milky Way galaxy itself will be
essential; the systematic uncertainty in the COBE quadrupole is about
70\% of its measured value \cite{K96}.  This very large uncertainty is
serious for measurement of the super-Hubble suppression scale.
On the one hand, this systematic uncertainty could fully account
for the suppressed quadrupole found by COBE and thus cast
doubt on the claimed measurement in \cite{bfh97}.  On the other hand,
this large uncertainty implies the size of the super-Hubble
suppression scale remains an open question and conceivably
it may be small enough to be detected. 

In this paper, we take an optimistic view that foreground subtraction
and parameter estimation (other than of the suppression scale) can be
performed precisely, and calculate the error with which the
suppression scale could be determined from CMB measurements.  We do
this via a likelihood analysis (Section \ref{Like}).  Section \ref{Open}
explores how an open inflation model, leading to $\Omega_0<1$, can
lead to a measurably small super-Hubble scale, which might provide
independent support for such a model.  In Section
\ref{ThCon} we explore the theoretical consequences of a super-Hubble
scale suppression, and draw our conclusions in section \ref{Concs}.

\section{$k_{\rm min}$ Super-Hubble detection: flat Universe}
\label{Like}

In this section we compute the errors expected on a measurement
of the suppression scale in a flat universe.  {}For a large
suppression scale, the signal will come from the low-order
multipoles in the CMB, which are dominated by the Sachs-Wolfe
effect.  The power spectrum of scalar temperature fluctuations is
$C_\ell \equiv \langle |a_\ell^m|^2\rangle$, where $a_\ell^m
\equiv \int\,d\Omega \Delta T(\Omega)/T$ and $T(\Omega)$ is the
temperature on the sky. {}For low $\ell$ the power spectrum may be
written in terms of the mass-density power spectrum $P(k)$ (e.g.
\cite{Efstathiou90}):
\begin{equation}
C_\ell = {4\pi\over 25} \int_0^\infty {dk\over k^2} j_\ell^2(\eta_0
k) P(k)
\end{equation}
where $j_\ell$ is a spherical Bessel function, and $\eta_0$ is the
coordinate distance of the last scattering surface, approximately
$2/H_0$ (aside from factors of $c$, we use dimensional units).

{}For the power spectrum, we assume minimal suppression as
suggested by \cite{zel65}, and a primordial power-law
spectrum:
\begin{equation}
P(k) \propto {k^n\over 1+\left({k\over k_0}\right)^{n-4}}.
\end{equation}
We assume $n=1$ in what follows.  \scite{rw96}
find that $k_0 \simeq 2.24/D$, where $D$ is the scale beyond
which there are no mass-density correlations.

Assuming all-sky coverage, estimates of $C_\ell$ are uncorrelated, and
we can compute the probability of $k_0$ given the data as a product of
individual likelihoods (assuming a uniform prior for $k_0$). To
estimate the error, and in this section only, we assume the likelihood
is gaussian (valid for high $\ell$, but not particularly accurate for
low multipoles), and compute the Fisher information matrix (scalar for
a single parameter) $F=-\langle \partial^2 \ln L/\partial
k_0^2\rangle$.  The error on $k_0$ is $\sigma_{k_0} = F^{-1/2}$.  Up
to an additive constant,
\begin{equation}
\ln L(k_0) \simeq -\sum_\ell \ln \sigma_\ell - \sum_\ell {(\hat C_\ell
- C_\ell(k_0))^2\over 2 \sigma_\ell^2}
\end{equation}
where $\hat C_\ell$ are the observed values, $C_\ell(k_0)$ is the
theoretical value for given $k_0$, and $\sigma_\ell^2 = 2C_\ell^2/(2\ell+1)$
is the variance assuming the temperature map is a gaussian random
field.  The error on $k_0$ from each multipole is then simply obtained 
from the Fisher matrix:
\begin{equation}
\sigma_{k_0} = {C_\ell \over \sqrt{\ell+1/2}\,|C'_\ell|}
\end{equation}
where $C'_\ell \equiv \partial C_\ell/\partial k_0$, and, in the case
of all-sky coverage, the estimators of $k_0$ may be close to
independent and can optimally be combined with inverse-variance weighting.
\begin{figure}
\begin{center}
\begin{picture}(200,150)
\includegraphics{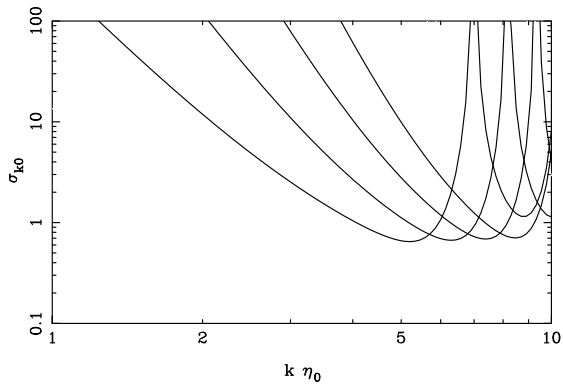}
\end{picture}
\end{center}
\caption[]{\label{Likeplot}
Fisher estimate of the error on $k_0$,
for individual multipoles from $\ell=2\ldots 5$, from left to right.
The model assumes a sharp cutoff in the power spectrum at
$k_0$. $\eta_0$ is the coordinate of the visible horizon. $\eta_0=2$
for an Einstein-de Sitter universe.}
\end{figure}
The contribution to the error from the low multipoles is illustrated
in Fig. \ref{Likeplot} for a model where we assume the limit of causal
processes introduces an abrupt cutoff in $P(k)$.  We see, as expected,
that the ability to detect the suppression scale is a very sensitive
function of the suppression scale itself.  There is a rather sharp threshold
of detectability at $k\eta_0 = 4$.  Note that for flat models
with positive $\Lambda$, the detection of a given physical suppression
scale is easier, since $\eta_0$ is larger than for an Einstein-de
Sitter model.  No analytic expression exists for $\eta_0$ in flat
models with non-zero $\Omega_\Lambda$ and $\Omega_0$: $\eta_0$ rises
from $2/H_0$ when $\Omega_\Lambda=0$ to $3.4/H_0$ when
$\Omega_\Lambda=0.7$.

This illustrative calculation is approximate because it assumes a
gaussian likelihood, and secondly, the likelihood surface is not
well-approximated by a gaussian near the peak.  Consequently, we
compute from now the full likelihood curve, assuming the proper
probability distribution for each multipole:
\begin{equation}
\ln L(k_0) = \sum_\ell \ln P\left(\hat C_\ell| C_\ell(k_0)\right)
\end{equation}
where $P(\hat C|C)$ is a $\chi^2$ distribution with $2\ell+1$
degrees of freedom.  Note, however, that the Fisher method,
which approximates $\ln(L)$ as parabolic around the peak, may not be
very accurate, since the likelihood is badly approximated by a
gaussian if the signal-to-noise is moderate, with a significant
probability tail extending to $k_0=0$.  This is explored further 
in section \ref{Open}.

\section{$k_{\rm min}$ Super-Hubble detection: Open Universe}
\label{Open}

This case is more interesting theoretically.  As the Universe
inflates, it is driven towards flatness.  In addition, the
particle horizon is inflated to larger physical scales.  It is
possible that inflation can produce a low-density universe, but
only with a degree of fine-tuning in current models.  Thus
observations implying a low density are considered problematic
for inflation.  However, the modest inflation of the particle
horizon offers the possibility of an independent check, as the
relatively small super-Hubble scale in this case may be
detectable.  In this section, we explore this possibility.

We follow in part the notation of \cite{Ellis88}, and make the following
assumptions: perturbations are generated by causal processes up
to the particle horizon size; the Universe is radiation-dominated
prior to inflation; the scale factor increases by a factor $10^Z$
during inflation, and by $10^{27-\Delta}$ after inflation.
$\Delta$ is set by the physics of inflation and for most models
lies near zero.  As argued in \cite{Ellis88}, such a model can account for
any present-day density parameter, depending on the pre-inflation
density parameter $\Omega_{i}$ and the amount of inflation:
\begin{equation}
1-\Omega_0 = E^2(1-\Omega_{i}) \label{EOmega}
\end{equation}
where $E = 10^{27-\Delta-Z}\Omega_{r0}^{1/2}$ is the fractional change
in $HR$ between the onset of inflation and the present day (assuming
zero cosmological constant after inflation).  $\Omega_{r0} = 4.22
\times 10^{-5} h^{-2}$ is the present-day radiation density parameter,
written in terms of the Hubble parameter $h \equiv H_0/(100 {\rm
km}\,{\rm s}^{-1}\,{\rm Mpc}^{-1})$.  Thus, for a given $\Omega_i$ and
particle physics model (via $\Delta$), (\ref{EOmega}) gives us the
present-day (open) density parameter in terms of the number of
inflation 10-foldings.

\begin{figure}
\begin{center}
\begin{picture}(200,150)
\includegraphics{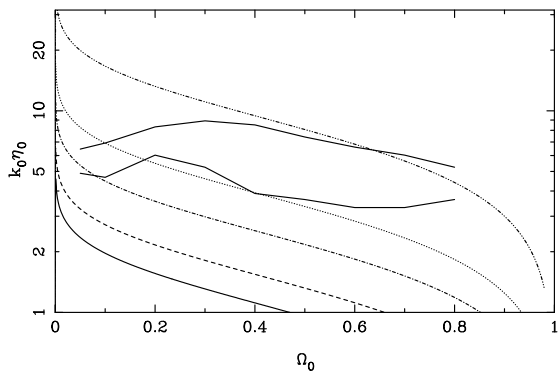}
\end{picture}
\end{center}
\caption[]{\label{k0eta0}
The suppression wavenumber $k_0$ in length
units of the distance to the last-scattering surface $\eta_0$, as a
function of the present density parameter, for a range of
pre-inflation density parameters, $\Omega_i$.  From the top,
$\Omega_i=$0.8, 0.3, 0.1, 0.01 and 0.001.  Also shown is the line
showing the minimum value of $k_0 \eta_0$ which can be distinguished
from $k_0=0$ for $n=1$ (top) and the Ratra-Peebles power spectrum
(bottom).  See text for details.}
\end{figure}

The suppression scale is also related to the number of
10-foldings:  the largest physical scale which is
causally-connected is the immediate pre-inflation particle
horizon, expanded to the present-day.  This scale is
\begin{equation}
D = {R(t_0)\over R(t_i)} \int_0^{t_i} {dt\over R(t)}
\end{equation}
which for radiation-domination prior to inflation, is
\begin{equation}
D = {E\over H_0} {\rm cosh}^{-1}\left(\Omega_{i}^{-1/2}\right).
\end{equation}
{}From \scite{rw96}, this gives a minimal suppression
wavenumber $k=2.24/D$, which we write in terms of the distance to the
last-scattering surface, $\eta_0$.  We approximate this by the
particle horizon assuming matter-domination throughout, $\eta_0 =
{\rm arccosh}(2/\Omega_0-1)$.
This is plotted in Fig. \ref{k0eta0}, for a range of pre-inflation density
parameters.

\begin{figure}
\begin{center}
\begin{picture}(200,150)
\includegraphics{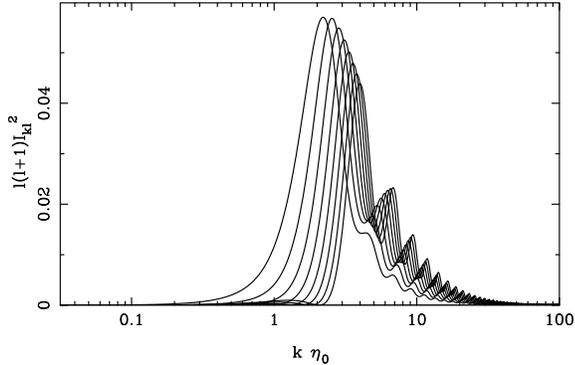}
\end{picture}
\end{center}
\caption[]{\label{Iplot}
Kernel functions giving the contribution to
multipoles $C_\ell;\ \ell=2,\ldots,9$ (from left to right) from power
at wavenumber $k$, for an Open Universe with $\Omega_0=0.2$.}
\end{figure}

The CMB power spectrum in open models is related to the matter
spatial power spectrum $P_R$ (defined in a non-standard way; $P_R$=
constant is a scale-invariant spectrum) by
\scite{LW95} (see also \pcite{GLLW95}, \scite{KS94}):
\begin{equation}
C_\ell(k_0) = 2 \pi^2 \int_0^\infty \,{dk\over k} P_R(k)I_{kl}^2
\end{equation}
The kernel $I_{kl}$ depends on a number of functions defined in
\scite{LW95} and \scite{GLLW95}, and includes Sachs-Wolfe and
Integrated Sachs-Wolfe effects.  We mention one practical issue: one
of these sets of functions ($\tilde\Pi_{kl}$) was computed using the
recurrence relation (15) in \cite{LW95}.  {}For small $r$ the recurrence
relation is unstable, and we instead used the asymptotic form
$\tilde\Pi_{kl}\rightarrow A_l(k)r^l$, computing the coefficients
using the recurrence relation in the limit $r\rightarrow 0$.  The
kernel functions $I_{kl}$ are shown in Fig. \ref{Iplot} for $l= 2$ to
9.  Note the presence of some supercurvature modes ($k<1$) in these
models.
\begin{figure}
\begin{center}
\begin{picture}(200,150)
\includegraphics{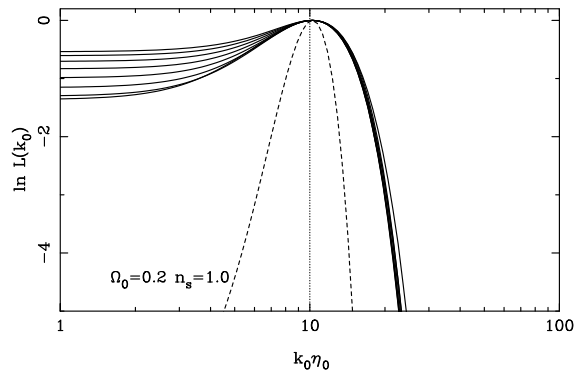}
\end{picture}
\end{center}
\caption[]{\label{LikeOpen}
Likelihood for the suppression wavenumber
$k_0$, from individual multipoles $C_\ell;\ \ell=2,\ldots,9$ (from
bottom to top), and the combined likelihood (dashed), from a suppressed
scale-invariant power spectrum, for an Open Universe with
$\Omega_0=0.2$.  The true $k_0$ is marked by the dotted line.  Clearly
in this case there is still some information from higher multipoles than
$\ell=9$.}

\end{figure}
In Fig. \ref{k0eta0} we also show the detectability limits for the
suppression wavenumber, for two choices of the primordial power
spectrum, a scale-invariant spectrum (top) and the power spectrum
proposed for open universes (bottom) by \scite{RatraPeebles}:
$P(q)\propto (4+q^2)^2/[|q|(1+q^2)]$ - see \scite{KS94} .  The lines
show the boundary above which the value of $k_0\eta_0$ for which the
relative likelihood for $k_0=0$ (compared with the true value) is less
than $e^{-4}=0.018$ (see Fig. \ref{LikeOpen} for an example of the
full likelihood curve).  This illustrates that, to a first
approximation, the detectability limit is determined by the present
conformal time: $\lambda \simeq \eta_0$.  We see that it is, in
principle, possible to detect the suppression scale in such an
inflationary model, provided $\Omega_i$ is large enough for a given
$\Omega_0$.  {}For example, we require $\Omega_i>0.3$ for $\Omega_0=0.1$
and a scale-invariant spectrum.  Note also that, roughly, the
detectability scale for the scale-invariant power spectrum is the
visible horizon size, i.e. $2\pi/k_0 \simeq \eta_0$; in particular, it
is possible to detect a cutoff on scales larger than the curvature
radius in low-density models.

\section{Theoretical Consequences}
\label{ThCon}

Any theoretical model has an implied super-Hubble scale.
In most models, it also is assumed that density perturbations are
produced from an initially smooth background, so that beyond the
super-Hubble scale the power is suppressed.  The analysis in
the earlier sections is applicable for such types of models.
It is worth mentioning that alternative possibilities are
conceivable beyond the super-Hubble scale, such as the opposite
case of a highly enhanced power spectrum and, in principle, any possibility
between these two limits.  Similar analyses as in the previous sections
could be done for other behaviour besides suppression beyond the
super-Hubble scale.  Nevertheless, the analysis in the previous sections
is more sensitive to detecting a disparate change in the power spectrum
at large scale rather than to the precise nature of this
disparity. Indeed, for large {\em enhancements} of power on
super-Hubble scales, detection is much easier \cite{GriZel78} 
Thus, we believe detection limits found in the earlier sections for
the super-Hubble scale are generic limits, fairly insensitive to
the nature of the power beyond this scale.  A conclusive test of this
belief would be to fully invert the power spectrum and determine
the intrinsic measurability limits imposed by cosmic variance, as suggested
by \cite{BM99}.  However, this will be left for future work.

Models of primordial density perturbations can be generally classified
as either inflationary or non-inflationary.  Inflationary models are
further differentiated between isentropic expansion which results in a
supercooled inflationary regime and non-isentropic expansion which
results in a warm inflationary regime.  {}For inflationary models,
primordial density perturbations are produced from quantum
fluctuations in supercooled inflation \cite{BST83,Hawking82,Starob82}
and thermal fluctuations in warm inflation \cite{BF95,Berera99}.  In
non-inflationary models, primordial density fluctuations emerge during
the radiation- and/or matter-dominated regimes from various possible
particle physics mechanisms such as cosmic strings
\cite{ASteb92a,ASteb92b} and late-time phase transitions
\cite{Wass86,PRS90,FS92,FHW92}.

In comparing supercooled inflation models to non-inflation models, the
prediction for the super-Hubble scale is distinctively different.  {}For
supercooled inflation models, the super-Hubble scale generically is
predicted to be many orders of magnitude larger than the Hubble radius
so that effectively $k_{\rm min} \approx 0$ (there are some
supercooled inflation models that are exceptions, in which a small
number of e-folds can occur \cite{Yoko99,KR99}).  In contrast, for
non-inflationary models, the causal horizon is about the same size as
the present-day Hubble radius, thus $k_{min} \sim H_0$ in these
models.  Specifically, in \scite{rw96} an analysis based on general
arguments found that the largest length scale on which sub-Hubble
scale perturbations can generate significant power is $\lambda
\stackrel{<}{\sim} 3\eta_0$ which corresponds to $k_{\rm min} \eta_0
\stackrel{>}{\sim} 2.2$.  Particular particle physics non-inflation
models of density perturbations are consistent with this general
estimate. {}For example cosmic strings \cite{ASteb92a,ASteb92b} plus
cold or hot dark matter find $k_{\rm min} \eta_0 \sim 2.1-7.9$ and
models of late-time phase transitions \cite{Wass86,PRS90,FS92,FHW92}
find $k_{\rm min} \eta_0 \gs 2.2$. 

The predictions for the super-Hubble scale in warm inflationary models
are intermediate to supercooled inflation and non-inflation models.
Unlike supercooled inflation models which generally favor a very
large number of e-foldings, phenomenological warm inflation
\cite{Berera97} and first-principles warm inflation
\cite{BGR99,Berera99} models are more democratic towards any number of
e-foldings.  One generic feature of warm inflation, that is of
potential interest for observation, is the number of e-folds of
inflation are correlated with the decrease in temperature of the
universe during the inflationary epoch \cite{Berera97}.  In distinct
contrast, such a correlation between the e-folds of inflation and the
temperature of the universe is not present in supercooled inflation.
Thus, if future data supports a small super-Hubble suppression scale,
within the measurability bounds, consistency of this finding with warm
inflation models could be checked through examining correlations with
the post-inflation temperature.  Such consistency checks would be much
less robust for supercooled inflation models.  Conversely, if
particular post-inflation temperatures are found by independent
theoretical and/or observational arguments, for warm inflation models,
such arguments also would imply predictions for the super-Hubble
scale.  In particular, arguments favoring higher post-inflation
temperatures, when applied to any given warm inflation model, also
would favor smaller number of e-folds and thus a smaller super-Hubble
scale.  {}Finally, in the warm inflationary era, isocurvature
perturbations also are generated due to thermal fluctuations in the
radiation field \cite{LF97,TB00}.  Since isocurvature modes do not
contribute to curvature perturbations (in the comoving gauge), the
contribution from isocurvature modes should be absent from
super-horizon perturbations, leading to a drop in amplitude of the
scalar perturbations at the horizon scale, thus imitating a
super-Hubble suppression scale.  Estimates of isocurvature
perturbations and their relative magnitudes compared to adiabatic
perturbations have been computed for a large class of warm inflation
models in \cite{TB00}.

In the open universe case, the commonly accepted particle physics
model for open-inflation is based on the Gott bubble nucleation
mechanism \cite{Gott82,GottStat84}.  In recent times, the
observational predictions from such models has been calculated to a
high degree of precision \cite{RatraPeebles,BGT95,YST95,STY95}.  In this
scenario there are two stages of inflation.  The first is a stage of
old inflation in which the universe is trapped in a metastable state.
Inside this sea of inflating false vacuum, a bubble is nucleated.  In
the scalar field description of this scenario, the bubble nucleation
process is described by an O(4) symmetric bounce solution
\cite{Coleman77,Coleman80}.  The key point is that the interior of the
bubble is a homogeneous and isotropic open universe, which solves the
horizon problem.  Inside the bubble, initially $\Omega_i$ is very
close to zero.  Then a second stage of inflation is pictured, this
time of the slow-roll new inflation type.  This inflation is fine
tuned to a small number of e-folds, just enough to achieve the desired
flatness of the universe with $\Omega_0 \stackrel{>}{\sim} 0.1$.

In this picture, the purpose of the first stage of old inflation
arises because bubbles typically nucleate at finite radius and
not at a point.  This requires that the universe should be smooth on the 
length scale of size the initial bubble when it
nucleates, and the first
epoch of old inflation achieves this smoothness.
The observable implications of this first stage
of inflation are primarily through
the initial fluctuations inside the bubble, and these have
been demonstrated to have minimal measurable
consequences in \cite{BGT95,YST95}.

In these open inflation models, the behavior of the background
cosmology is the same as the treatment in \cite{Ellis88} and applied
in Sect. \ref{Open}, to the extent of relating $\Omega_0$, $\Omega_i$ and the
number of e-folds of inflation.  We have not attempted a detailed
analysis of these models, but some general remarks are appropriate.
In these open inflation models, the relation of these quantities to
the super-Hubble scale is not the same as in Sect. \ref{Open}, since here two
stages of inflation occur.  In these models, the entire visible
universe sits inside a nucleated bubble and this bubble sits inside a
larger universe which was created by the first stage of old inflation.
The super-Hubble scale contains this entire system
and so is very large, $k_{min} \approx 0$, thus unmeasurable.
However these models have an intermediate scale smaller than this
super-Hubble scale and larger than the Hubble radius inside the
nucleated bubble, and this is the distance to the bubble wall.  There
clearly are fluctuations on this scale \cite{YST96}, 
and this might be measurable
by the methods of this paper.
{}From our analysis in Sect. \ref{Open}, it can be inferred
that any effect just slightly larger than the Hubble radius could be
detected.  In this respect, these open universe models do not rule out
the possibility that the distance to the bubble wall may be just a
little larger than the curvature scale. In this case, in the low
density regime, where the curvature scale and the Hubble radius are
about the same size, the bubble wall will create some sort of
super-Hubble effect that is not too far outside the Hubble radius.
Although we have not considered any detailed model for this situation,
from the analysis in Sect. \ref{Open} it can be inferred that such effects may
be measurable.


\section{Conclusion}
\label{Concs}

We have investigated the extent to which a large-scale suppression
scale in primordial fluctuation power might be detectable in flat and
open universe models.  We have concentrated on observations of the
microwave background radiation, as direct measurement of the
suppression scale from the galaxy power spectrum is likely to be
impossible, since the scale is likely to be at least a substantial
fraction of the Hubble radius.  The suppression itself is expected if
the fluctuations arise from causal processes.  The limit of
detectability is dependent a little on cosmology, and on the
primordial power spectrum, but it is no surprise that it is set
roughly by the size of the visible horizon today.  In flat models, a
positive cosmological constant makes the suppression scale easier to
detect, as the visible horizon is larger.  Defect models will be
distinguished from inflationary models most easily by their
high-$\ell$ power spectrum or by non-gaussian tests
(e.g. \pcite{HS99}).  Should inflation models be ruled out, then the
suppression scale may become a useful discriminant between defect
models, as it should appear on scales smaller than the visible
horizon, and be readily detectable.  Clearly, it would also act as a
useful cross-check.  

{}For single-inflation models, the suppression
scale depends on both the final value of the matter density parameter,
and also on the radiation density parameter immediately prior to
inflation. {}For low final densities, there is the possibility of
detecting a moderately small suppression scale, which would provide
some sort of independent evidence on what would be a fine-tuned model.
{}For open inflation models involving two stages of inflation, there
appears to be no possibility of detection of the causal suppression
scale set by the first stage of old inflation.  However, as mentioned
in Sect. \ref{ThCon}, we can infer 
from the analysis in Sect. \ref{Open} that any
significant effect that is slightly super-Hubble scale could be
detectable.  In particular, if the bubble wall created in the second
stage of inflation is just beyond the Hubble radius, the type of
measurements we have been considering could be applied to detect
it. However, we have not made any detailed analysis of this
possibility.  

We have considered only scalar modes here; in principle
it is possible to include tensor modes to increase the detectability,
but in fact the transition between a scale which is easily detectable
to one which is not is so sharp that such additional information would
change the results very little.  We expect the suppression scale
signal to be contained in the low-order multipoles, which were
measured by COBE.  {}Future experiments such as MAP and Planck will
improve on COBE not principally because of better signal-to-noise, but
via improved foreground subtraction through their better frequency
coverage.  In addition, their better resolution will allow better
determination of cosmological parameters.  The reduced uncertainty in
other parameters will decrease the marginal error on the suppression
scale.  This is quantifiable in principle, but the number of
cosmological parameters is large (up to $\sim 20$ \scite{BET97}).  It
is in the better foreground subtraction that MAP and Planck should
gain: the systematic uncertainty in the COBE quadrupole is comparable
to the measured signal \cite{K96}, large enough that the apparent
suppression of the quadrupole in the COBE data \cite{bfh97} could be
entirely due to this rather than super-Hubble suppression.  Provided
that foregrounds can be adequately removed and other cosmological
parameters are determined accurately from the high-order
power spectrum, a factor of 3 improvement
in the quadrupole noise is expected, and this should allow the
super-Hubble scale suppression to be estimated to the accuracy
presented in this paper.  If such a scale is indeed measured, then, as
with other measurements, polarization measurements can be used as a
consistency check.

\begin{acknowledgments}
We are particularly grateful to Bharat Ratra for many useful
discussions.  Computations were made using STARLINK facilities. AB is
funded by PPARC.
\end{acknowledgments}


\end{document}